\documentclass[a4paper,11pt]{article}
\pdfoutput=1 % if your are submitting a pdflatex (i.e. if you have
\usepackage{jheppub} % for details on the use of the package, please
\usepackage[T1]{fontenc} % if needed

\title{\boldmath Constraints on Dark Energy state equation with varying pivoting
redshift}

\author[a,1]{Dario Scovacricchi,\note{Corresponding author.}}
\author[a,b,c]{Silvio A. Bonometto}
\author[a,b]{Marino Mezzetti}
\author[d,e]{Giuseppe La Vacca}

\affiliation[a]{Trieste University, Physics Department, Astronomy Unit, Via Tiepolo 11--13, 34143 Trieste, Italy}
\affiliation[b]{I.N.A.F. Trieste Astronomical Observatory}
\affiliation[c]{I.N.F.N. Sezione di Trieste}
\affiliation[d]{Milano--Bicocca University, Physics Department, Piazza
  della Scienza 3, 20126 Milano, Italy}
\affiliation[e]{I.N.F.N. Sezione di Milano--Bicocca}

\emailAdd{dario.scovacricchi@port.ac.uk}

\abstract{We assume the DE state equations $w(a) = w_0+w_a(a_p-a) $,
  and study the dependence of the constraints on $w_0$ and $w_a$
  coefficients on the pivoting redshift $1+z_p=1/a_p$. Coefficients
  are fitted to data including WMAP7, SNIa (Union 2.1), BAO's
  (including WiggleZ and SDSS results) and $H_0$ constraints. The
  fitting algorithm is CosmoMC. We find specific differences between
  the cases when $\nu$--mass is allowed or disregarded. More in
  detail: (i) The $z_p$ value yielding uncorrelated constraints on
  $w_0$ and $w_a$ is different in the two cases, holding $\sim 0.25$
  and $\sim 0.35$, respectively. (ii) If we consider the intervals
  allowed to $w_0$, we find that they shift when $z_p$ increases, in
  opposite directions for vanishing or allowed $\nu$--mass. This leads
  to no overlap between 1$\sigma$ intervals already at $z_p >\sim
  0.4$. (iii) The known effect that a more negative state parameter is
  required to allow for $\nu$ mass displays its effects on $w_a$,
  rather than on $w_0$. (iv) The $w_0$--$w_a$ constraints found by
  using any pivot $z_p$ can be translated into constraints holding at
  a specific $z_p$ value (0 or the $z_p$ where errors are
  uncorrelated). When we do so, error ellipses exhibit a
  satisfactory overlap.
}

\begin{document} 
\maketitle
\flushbottom

\section{Introduction}
Owing to the conceptual problems of $\Lambda$CDM, a number of options
for Dark Energy (DE) nature have been considered. In particular, DE
could be a scalar field, necessarily self--interacting and possibly
interacting with Dark Matter
\citep{coupledquintessence1,coupledquintessence2,coupledquintessence3,
  coupledquintessence4,coupledquintessence5,coupledquintessence6,
  coupledquintessence7,coupledquintessence7bis,coupledquintessence8,
  coupledquintessence9,coupledquintessence10,coupledquintessence12,
  coupledquintessence13,coupledquintessence14,coupledquintessence15,
  coupledquintessence16,coupledquintessence17,coupledquintessence18,
  coupledquintessence19}, or just a phenomenological consequence of
large scale GR violations
\citep{GRviolate1,GRviolate2,GRviolate4,GRviolate5,GRviolate6}. But
neither these options, nor still more exhotic hypotheses
\citep{void1,void2,void3,void4,void5}, led to appreciable improvements
of the fit between theory and
data~\citep{coupledfit1,coupledfit2,coupledfit3}.

The problem has then been tackled from the phenomenological side, by
testing whether any linear $w(a)$, different from $w(a) \equiv -1$,
improves data fits. A possible option amounts then to express the
linear laws through the equations
\begin{equation}
\label{SE1}
w(a) = w_0 + w_a (1-a)~,
\end{equation}
aiming then at testing how various sets of data yield constraints on
$w_0$ and $w_a$.  Here $a$ is the scale factor, normalized to unity at
the present time. In the literature, this expression for $w(a)$ was
first used by \cite{chevall}. 

The same linear laws can be expressed also through the
equations
\begin{equation}
w(a) = w_{0,a_p} + w_{a,a_p} (a_p-a)
\label{SE}
\end{equation}
which differ from (\ref{SE1}) for selecting a non--vanishing pivoting
redshift
\begin{equation}
z_p = 1/a_p - 1~,
\label{pivot}
\end{equation}
while we put an extra index to the linear coefficients $w_{0,a_p}$,~$
w_{a,a_p} $ to put in evidence that, when changing $z_p$, their
values change. The straight lines defined by eq.~(\ref{SE1}) and
eq.~(\ref{SE}) are however the same: any equation (\ref{SE}) turns
into an equation (\ref{SE1}) if we set
\begin{equation}
w_{0,a_p} = w_0-w_a(a_p-1)
\end{equation}
and $w_{a,a_p}=w_a$. Notice that this last identity does not imply
that limits on $w_{a,a_p}$ are independent from the pivoting redshift.
In the sequel, whenever this causes no confusion, we shall however
follow the common use and call $w_0$,~$w_a$ the two parameters in any
expression (\ref{SE}).

Linear laws can be fitted to data by using different $a_p$ values.
Here we aim at testing, first of all, how compatible are results
obtained when varying the pivoting redshift.

We shall do so in two cases: either neglecting or allowing the option
that $M_\nu = \sum_\nu m_\nu \neq 0$ (the sum is extended to the mass
eigenvalues for 3 standard neutrino flavors). Let us also remind that
the neutrino density parameter
\begin{equation}
\Omega_\nu h^2 = 1.08 \times 10^{-2} (M_\nu/{\rm eV})
(T_{0\gamma}/2.73\, \, {\rm K})^3~,
\end{equation}
so that, when the dark matter reduced density parameter $\omega_c =
\Omega_c h^2$ is assigned, the neutrino fraction $f_\nu =
\Omega_\nu/\Omega_c$ immediately follows.

The value of $a_p$ can be selected so to have uncorrelated
phenomenological constraints on $w_0$ and $w_a$. Here we also
wish to put in evidence that: (i) the pivoting redshift yielding
uncorrelated constraints is different, if $f_\nu \equiv 0$ or can be
$\neq 0$; (ii) also the dependence on $a_p$ of the $w_0$ interval
compatible with data depends on the above option.

%%%%%%%% 

We expect that the DE state parameter $w$ takes lower values, even in
the phantom domain, when $M_\nu$ is allowed. We shall test how this
occurs, when we consider a wide set of data (see below). In
particular, by allowing for (linearly) variable $w$, we can test
whether data require a constantly low $w_0$ or a progressively
decreasing law, set by a negative~$w_a$.

%%%%%%%%%%%%%%%%%%%%%%%%%%%%%%%%%%%%%%%%%%%%%%%%%%%%%%%%%%%%%%%%%%%
\begin{figure}
\centering
\includegraphics[height=6cm,angle=0]{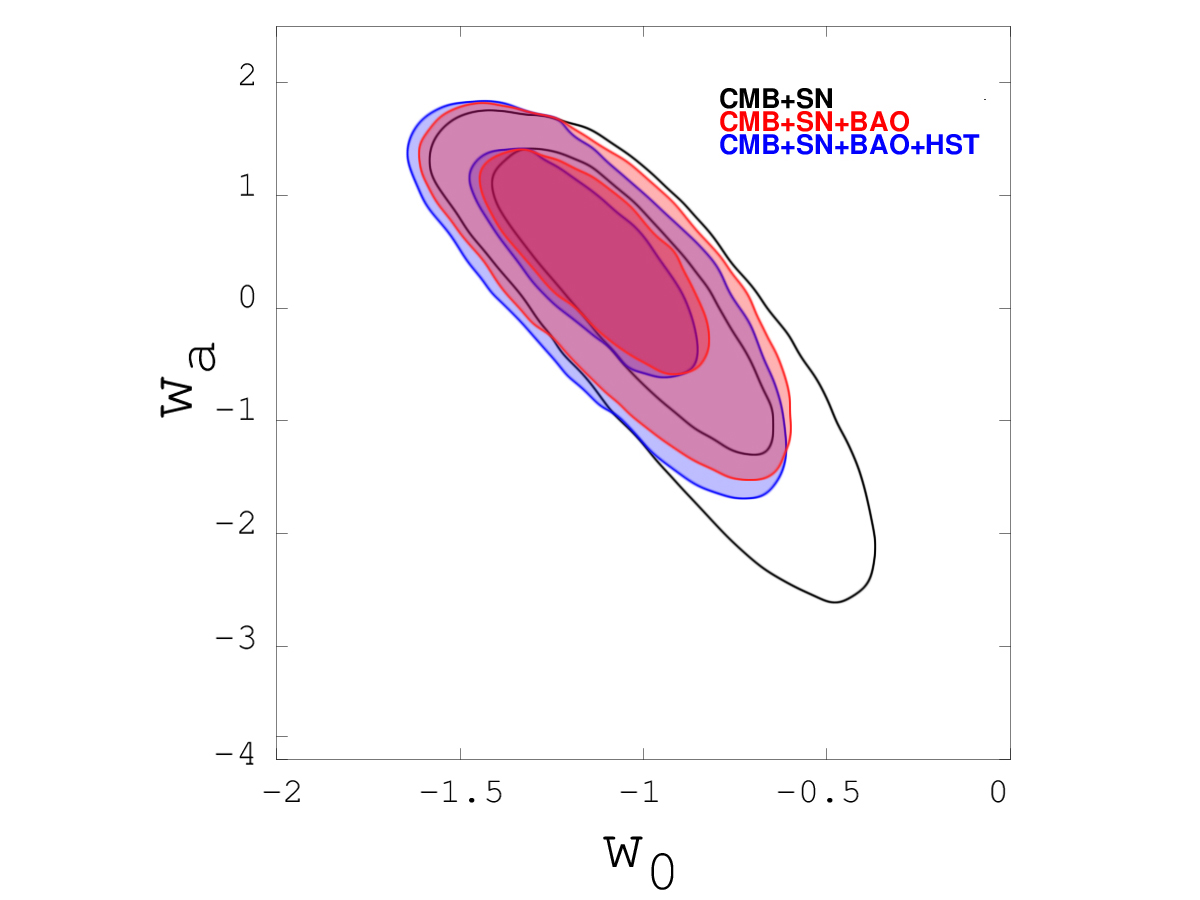}
\includegraphics[height=6cm,angle=0]{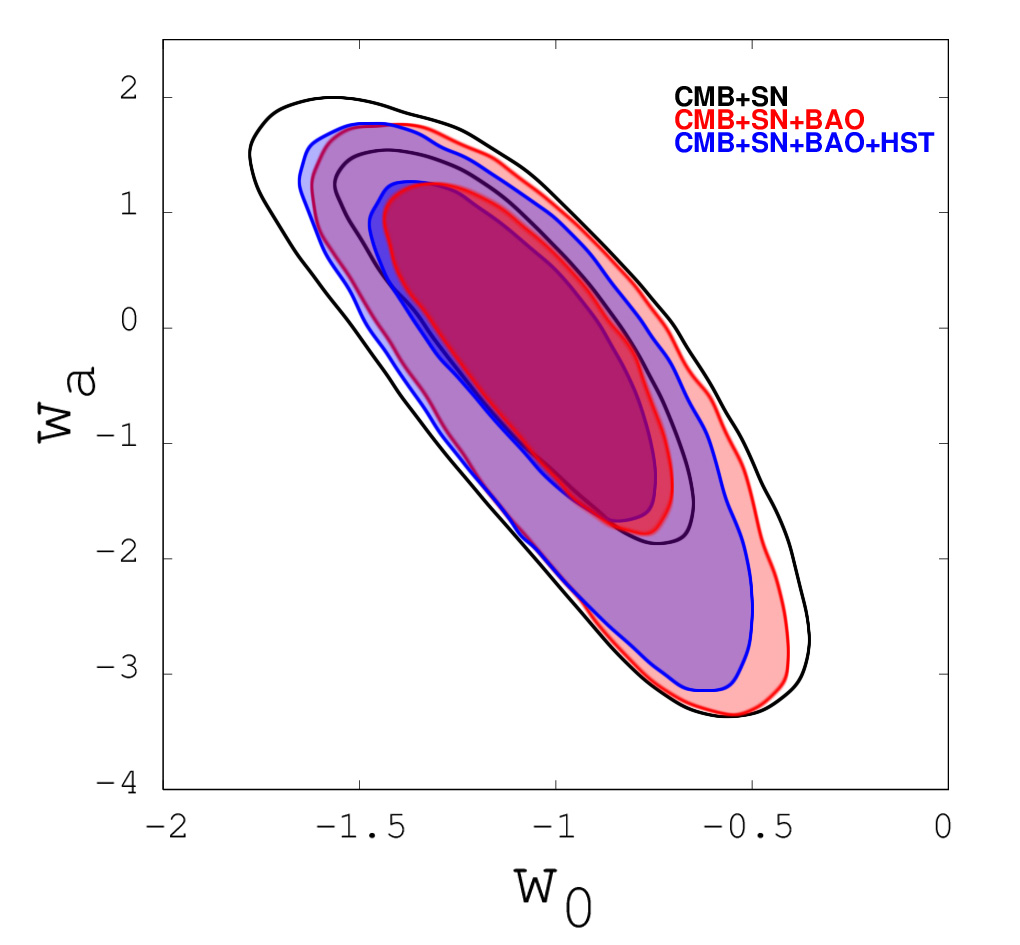}
\vskip 0.5truecm
\caption{2D marginalized likelihood, when taking $z_p=0$, on the plane
  $w_0-w_a$ at 65\% and 95\% of confidence for different data sets,
  when assuming massless (l.h.s.) or massive (r.h.s.) neutrinos.  The
  coordinate scale being the same on both sides allows us to
  appreciate how the $w_0$--$w_a$ uncertainty increases when a
  $\nu$--mass degree of freedom is considered. }
\label{fig1}
\vskip .5truecm
\end{figure}
%%%%%%%%%%%%%%%%%%%%%%%%%%%%%%%%%%%%%%%%%%%%%%%%%%%%%%%%%%%%%%%%%%%

In the recent literature, the set of linear $w(a)$ has also been
parametrized by using the values taken by $w$ at $z=0$ and at a higher
redshift, e.g. $z=0.5~.$ In spite of advantages of this
parametrization (\cite{wang}), quite a few authors still keep to the old
one. We plan to deepen the relation with such approach in further
work.

%%%%%%%%%%%%%%%%%%%%%%%%%%%%%%%%%%%%%%%%%%%%%%%%%%%%%%%%%%%%%%%%%%%%

\section{Results for $z_p = 0$.}
Let us then report, first of all, the results of Monte Carlo fits of
DE state equations {\it vs.}~data, performed by using the algorithm
CosmoMC\footnote{ http://www.cosmologist.info/cosmomc}
(\cite{cosmomc}, May 2010 version); the CosmoMC code was integrated
with the first version of the PPF module\footnote{camb.info/ppf} for
CAMB\footnote{http://www.camb.info/} (\cite{camb}, \cite{ppf}). Fits
were performed in respect to the following parameters: $w_0$, $w_a$
(in eq.~\ref{SE}) and $\omega_b = \Omega_b h^2$, $\omega_c = \Omega_c
h^2$, $\theta = 100~ l_s/l_d$, $\tau$, $n_s$, $\log A$, $A_{SZ}$, plus
$f_\nu$ when needed (respectively: reduced baryon density parameter,
reduced CDM density parameter, 100 times the ratio between sound
horizon at recombination and its angular diameter distance, optical
depth due to reionization, primeval spectral index, logarithmic
fluctuation amplitude with pivoting scale $0.05\, \, $Mpc$^{-1}$, SZ
template normalization, neutrino fraction as defined below; $h$ is the
Hubble parameter in units of 100 (km/s)/Mpc). We however kept
$\Omega_k=0$.

Our data set includes CMB data from WMAP7\footnote{Provided by the
  website lambda.gsfc.nasa.gov}, supernovae from Union2.1 survey
(\cite{union21}, option with no systematic errors), WiggleZ and SDSS
BAO's data (\cite{wigglez}, \cite{sdss}), HST data (\cite{hst}) and
CMB lensing as provided by CosmoMC.  We use different combinations of
these data, as suitably detailed below.

%In Figure \ref{w0wa_mass2} we show $1\sigma$ and $2\sigma$ contours
In Figure 1 we show $1\sigma$ and $2\sigma$ contours
for the marginalized likelihood on the $w_0$--$w_a$ plane, when $z_p =
0$, for the sets of data indicated in the frame. In comparison with
the analogous curves shown in WMAP7 report \cite{wmap7} for $f_\nu=0$,
our ellipses are slightly displaced towards more negative $w_0$
and greater $w_a$. The ranges found are closer to the Union 2.1 report
by \cite{union21}.

To gauge the widening of $w_0$ and $w_a$ intervals when $M_\nu \neq 0$
is allowed, we kept the same abscissa and ordinate ranges in both
sides. The widening is confirmed when fully marginalizing in respect
to all other parameters, as is shown in Table 1. Let us however notice
that, when $M_\nu \equiv 0$ is required, the inclusion of BAO and/or
HST data in top of CMB, causes a displacement towards smaller values
of the mean $w_0$ and an increase of $w_a$. These shifts --just below
$1\sigma$-- do not occur (or are much smaller) when $M_\nu \neq 0$ .

%%%%%%%%%%%%%%%%%%%%%%%%%%%%%%%%%%%%%%%%%%%%
\begin{table}
%\vskip -2truecm
\hspace{-6.truecm}

 \begin{tabular}{p{0.12\textwidth}|p{0.155\textwidth}p{0.155\textwidth}|p{0.155\textwidth}p{0.155\textwidth}p{0.16\textwidth}}

\hline & \hfill Massless &Neutrinos\hfill & \hfill Massive&
Neutrinos\hfill&\\ \hline \hline Data Set &$w_0\pm1\sigma\pm2\sigma$
&$w_a\pm1\sigma\pm2\sigma$ & $w_0\pm1\sigma\pm2\sigma$ &
$w_a\pm1\sigma\pm2\sigma$&$f_\nu\pm1\sigma\pm2\sigma$ \\ \hline \hline
CMB+SN & $-0.99_{-0.13-0.36}^{+0.11+0.41}$ &
$-0.04_{-0.36-1.74}^{+0.56+1.25}$&$-1.06_{-0.14-0.45}^{+0.13+0.46}$
&$-0.25_{-0.44-2.14}^{+0.68+1.56}$ &
$0.045_{-0.045-0.045}^{+0.013+0.052}$ \\ ....+BAO &
$-1.10_{-0.10-0.30}^{+0.08+0.33}$ &$~~~0.33_{-0.24-1.25}^{+0.37+0.92}$ &
$-1.03_{-0.12-0.35}^{+0.10+0.39}$ &$-0.37_{-0.36-2.01}^{+0.60+1.41}$
&$0.040_{-0.040-0.040}^{+0.011+0.043}$ \\ ....+HST &
$-1.13_{-0.10-0.30}^{+0.09+0.33}$ & $~~~0.29_{-0.25-1.30}^{+0.39+1.00}$
&$-1.07_{-0.11-0.34}^{+0.10+0.38}$ &$-0.31_{-0.34-1.89}^{+0.56+1.37}$
&$0.036_{-0.036-0.036}^{+0.009+0.041}$ \\ \hline

\end{tabular}
\label{tavola1}

\caption{Mean, $1\sigma$ limits and $2\sigma$ limits on $w_0$, $w_a$
  and $f_\nu$ (when relevant), with and without massive neutrinos and
  different data sets. Values are obtained by fully marginalizing over
  all other parameters. As indicated, in the second line BAO data are
  added to CMB+SN data, in the third line also HST data are included.
  The pivoting redshift is however $z_p = 0$.
}

\vskip .5truecm
\end{table}
%%%%%%%%%%%%%%%%%%%%%%%%%%%%%%%%%%%%%%%%%%%%%%%%%%%%%%%%%%%%%%%%%%%%%%%%%%%

%%%%%%%%%%%%%%%%%%%%%%%%%%%%%%%%%%%%%%%%%%%%%%%%%%%%%%%%%%%%%%%%%%%%%
%%%%%%%%%%%%%%%%%%%%%%%%%%%%%%%%%%%%%%%%%%%%%%%%%%%%%%%%%%%%%%%%%%%
\begin{figure}
\centering
\includegraphics[height=6cm,angle=0]{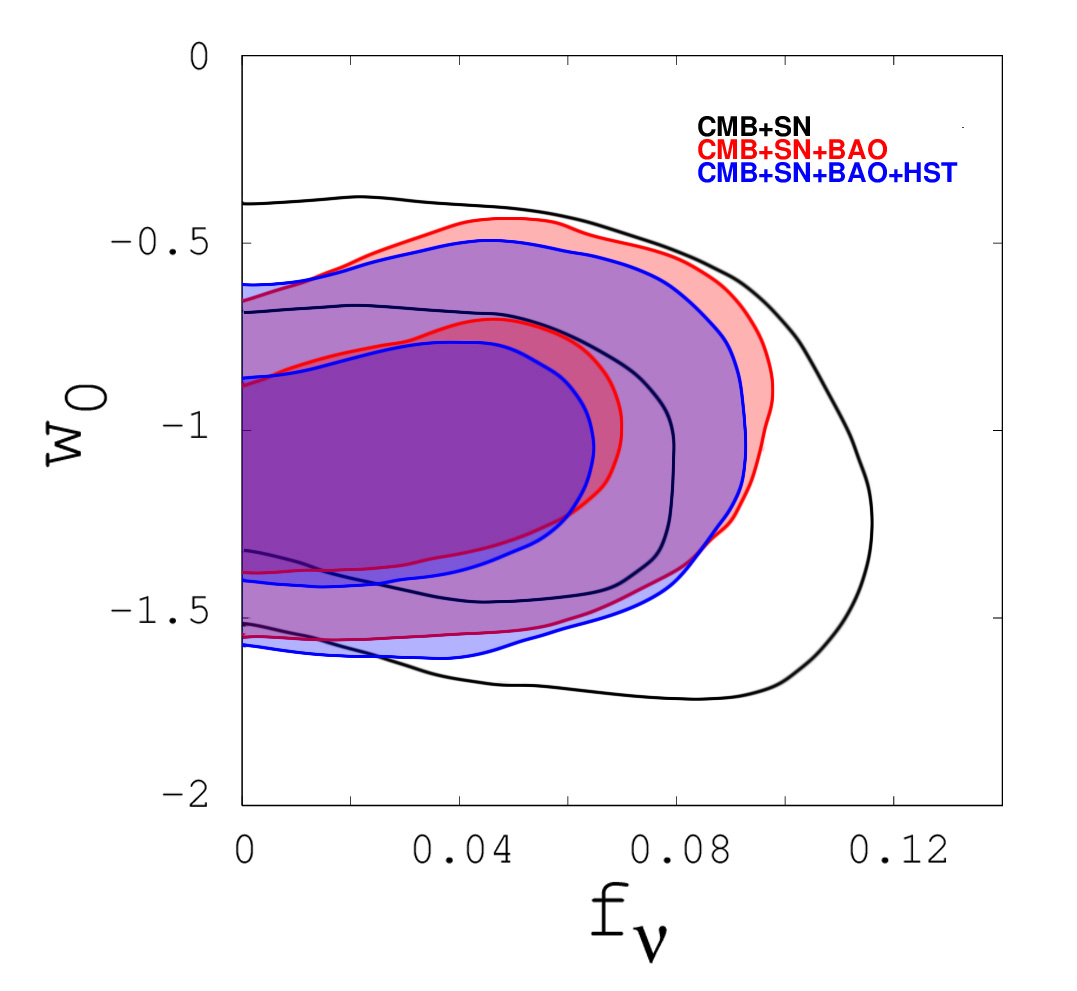}
\includegraphics[height=6cm,angle=0]{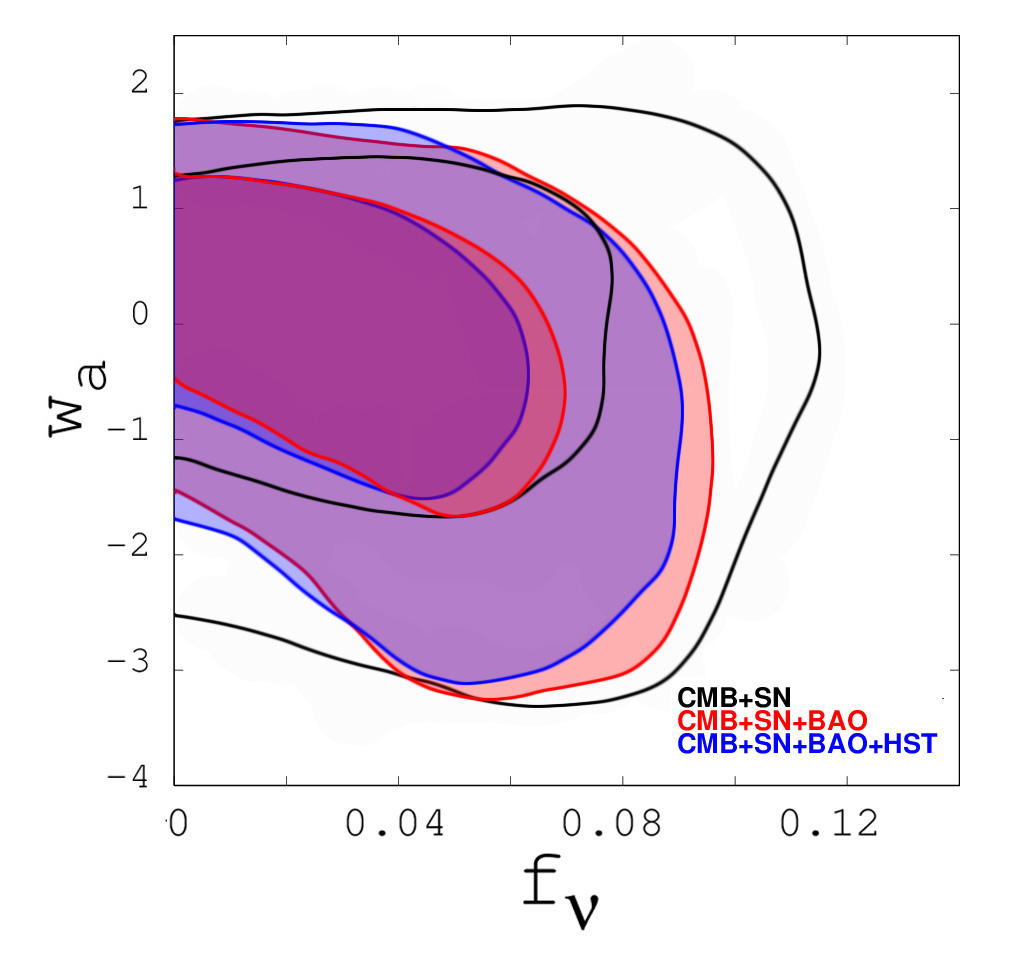}
\vskip 0.5truecm
\caption{2D marginalized likelihood on the plane $f_\nu-w_0$ (l.h.s.)
  and $f_\nu-w_a$ (r.h.s.) at 65\% and 95\% of confidence for
  different data sets, obtained with $z_p=0$.}
\label{fig2}
\vskip .5truecm
\end{figure}
%%%%%%%%%%%%%%%%%%%%%%%%%%%%%%%%%%%%%%%%%%%%%%%%%%%%%%%%%%%%%%%%%%%

%In Figure \ref{fnuw0_massive} we show the likelihood distributions on
In Figure 2  we show the likelihood distributions on
the $f_\nu$--$w_0$ and $f_\nu$--$w_a$ planes, outlining a progressive
delving of $w_a$ into the negative domain when $f_\nu$ shifts from 0
to 0.04~ (i.e., when $M_\nu$ shifts from 0 to $\sim 0.60~$eV). The
known result that a greater $M_{\nu}$ is allowed, when $w$ delves in
the phantom area, therefore affects $w_a$ rather than $w_0$, so
indicating that, to soften $M_{\nu}$ limits, it seems preferable that
$w$ shifts below -1 just when $z>0~.$

More in detail, if constant $w$ models only are considered, as in
\cite{wmap7} or \cite{phantom}, one finds that, to compensate a
massive neutrino component, DE density fading more rapidly than in
$\Lambda$CDM, as $z$ increases, is favored. As is known, even for
$M_\nu \sim 0.1\, $eV, neutrino derelativization is complete before $z
= 100~.$ Since then, the whole linear fluctuation spectrum evolves $
\propto (1+z)^{-1}$ until the spectral growth is slowed down by DE
acquiring a significant density. If $w$ is constant and $ < -1$, DE
density becomes significant later than in $\Lambda$CDM; a later slow
down compensates the spectral depression, which is one of the
consequences of neutrino mass. For instance, \cite{phantom} found a
constant $w=-1.12 \pm 0.09$.  With our wider dataset, we found
$w=-1.11^{+0.05}_{-0.04}$, an almost coincident result, apart of a
halvened (1$\sigma$) errorbar.

When $w$ linear variations are allowed, errors become greater, as
expected. The central point of $w_0$ ($\simeq -1.07$) however rises up
to the 1$\sigma$ limit for constant--$w$ models, while $w$ tends to
become negative because of $w_a.$ This means a different timing in the
reduced slowing down. A natural guess is that data coming from the
epoch when DE starts to become significant, e.g. WiggleZ data, are
better fitted by an early spectrum not only higher than $\Lambda$CDM,
but even higher than a $w \simeq -1.12$ phantom model.

\section{Results for $z_p \neq 0$}
Let us then consider the fits when pivoting redshifts $z_p \neq 0$ are
%considered. In Figure \ref{w0wa_pivot_massless} we overlap the
considered. In Figure 3  we overlap the
2$\sigma$ contour ellipses on the $w_0$--$w_a$ plane, for $z_p
= 0$, 0.25, 0.35 and 0.5, both for $f_\nu=0$ (l.h.s.) and $\neq 0$
(r.h.s.). For the sake of clarity, we consider only the full set of
observational constraints (CMB+SN+BAO+HST). The ellipses exhibit
a progressive straightening of the symmetry axes and $w_0$--$w_a$
errors become uncorrelated when $z_p \simeq 0.35$ or $z_p \simeq
0.25$, in the cases $f_\nu=0$ or $\neq 0$. More precisely, with the
system of data used here, the covariance Cov$(w_0,w_a) = \langle (w_0
- \bar w_0) (w_a - \bar w_a) \rangle$ vanishes for $z_p = 0.33$ or
%$z_p = 0.24$, respectively, as is shown by Figure \ref{contiboth}.
$z_p = 0.24$, respectively, as is shown by Figure 4.
Here $\bar w_0$ and $\bar w_a$ are mean values at those redshifts.

%%%%%%%%%%%%%%%%%%%%%%%%%%%%%%%%%%%%%%%%%%%%%%%%%%%%%%%%%%%%%%%%%%%
\begin{figure}
\centering
\includegraphics[height=8cm,angle=0]{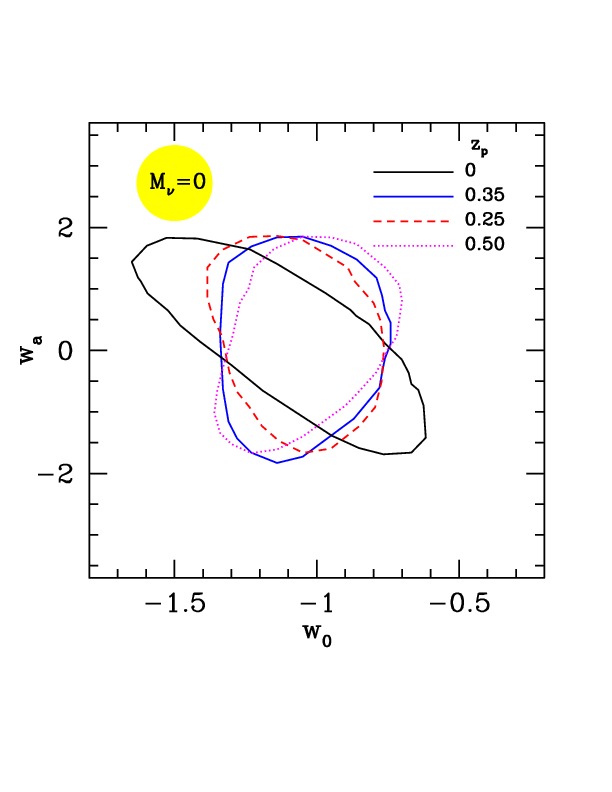}
\includegraphics[height=8cm,angle=0]{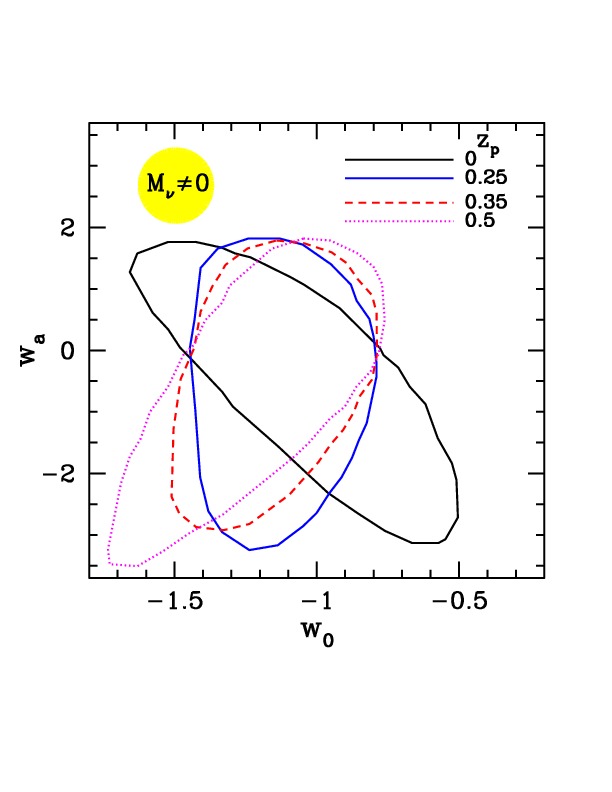}
\vskip -0.5truecm
\caption{2$\sigma$ marginalized likelihood contours on the $w_0-w_a$
  plane for different values of $z_p$, as indicated in the frame;
  $w_0$ and $w_a$ exhibit uncorrelated errors for $z_p \simeq 0.35$ (0.25)
  when $f_\nu \equiv 0$ ($\neq 0$). Notice the color and linetype
  inversion between $z_p = 0.25$ and 0.35, in the two
  Figures. All plots refers to CMB+SN+BAO+HST constraint combination.
}
\label{fig3}
\vskip .5truecm
\end{figure}
%%%%%%%%%%%%%%%%%%%%%%%%%%%%%%%%%%%%%%%%%%%%%%%%%%%%%%%%%%%%%%%%%%%

%%%%%%%%%%%%%%%%%%%%%%%%%%%%%%%%%%%%%%%%%%%%%%%%%%%%%%%%%%%%%%%%%%%
\begin{figure}
\centering
\includegraphics[height=10cm,angle=0]{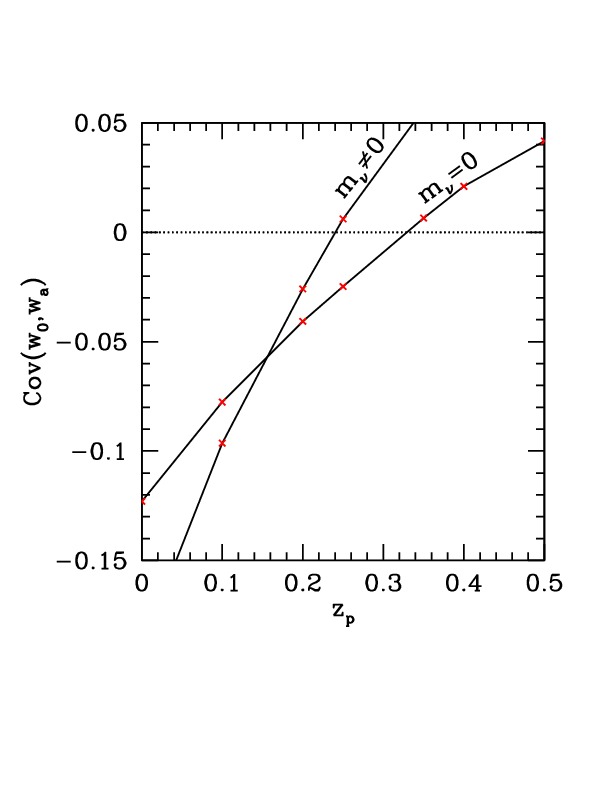}
\vskip -1truecm
\caption{Dependence on $z_p$ of the $w_0$--$w_a$ covariance.  This
  plot confirms that the two parameter estimates are statistically
  independent for $z_p \simeq 0.35~\, (0.25)$ if neglecting (allowing)
  neutrino mass. According to the plot, statistical independence
  occurs exactly at $z_p = 0.33 ~\, (0.24)$, respectively. Data
  improvements might however cause small shift of these ``exact''
  values.}
\label{fig4}
\vskip 0.5truecm
\end{figure}
%%%%%%%%%%%%%%%%%%%%%%%%%%%%%%%%%%%%%%%%%%%%%%%%%%%%%%%%%%%%%%%%%%%

Such difference between the two cases is a result of this analysis.
Notice again that the overall ellypso\"\i dal areas are much greater
on the r.h.s. This is the effect of adding just one extra parameter
and confirms that a significant correlation exists between the allowed
$w_0$--$w_a$ domains and $M_\nu$, so that neglecting the $\nu$--mass
option can be badly misleading, when we aim to constrain the DE state
equation.

%The ellipses in Figure \ref{w0wa_pivot_massless} apparently undergo a
The ellipses in Figure 3 apparently undergo a
gradual distortion and migrate through the plot. This is because each
coefficient pair $w_0$--$w_a$ (i.e. $w_{0,a_p}$--$w_{a,a_p}$)
corresponds to a different straight line, when $a_p$ varies.

It is however possible to translate the constraints found at any $a_p$
into constraints on the $w_{0,a_p=1}$--$w_{a,a_p=1}$ plane, or 
into constraints on the plane spanned by the coefficients
$w_{0}$--$w_{a}$ when $z_p = 0$, or when $z_p$ yields uncorrelated
%errors.  In the Figures \ref{overlaps} we show the shapes of the
errors.  In the Figures 5  we show the shapes of the
ellipses after this transformation. The two Figures on the first
(second) line refer to $f_\nu \equiv 0$ ($f_\nu \neq 0$). The Figures
at the l.h.s. (r.h.s.) are the ellipses at $z=0$ ($z=0.35$ or
0.25 close to where parameter errors are uncorrelated).

%%%%%%%%%%%%%%%%%%%%%%%%%%%%%%%%%%%%%%%%%%%%%%%%%%%%%%%%%%%%%%%%%%%
\begin{figure}
\centering
\includegraphics[height=8cm,angle=0]{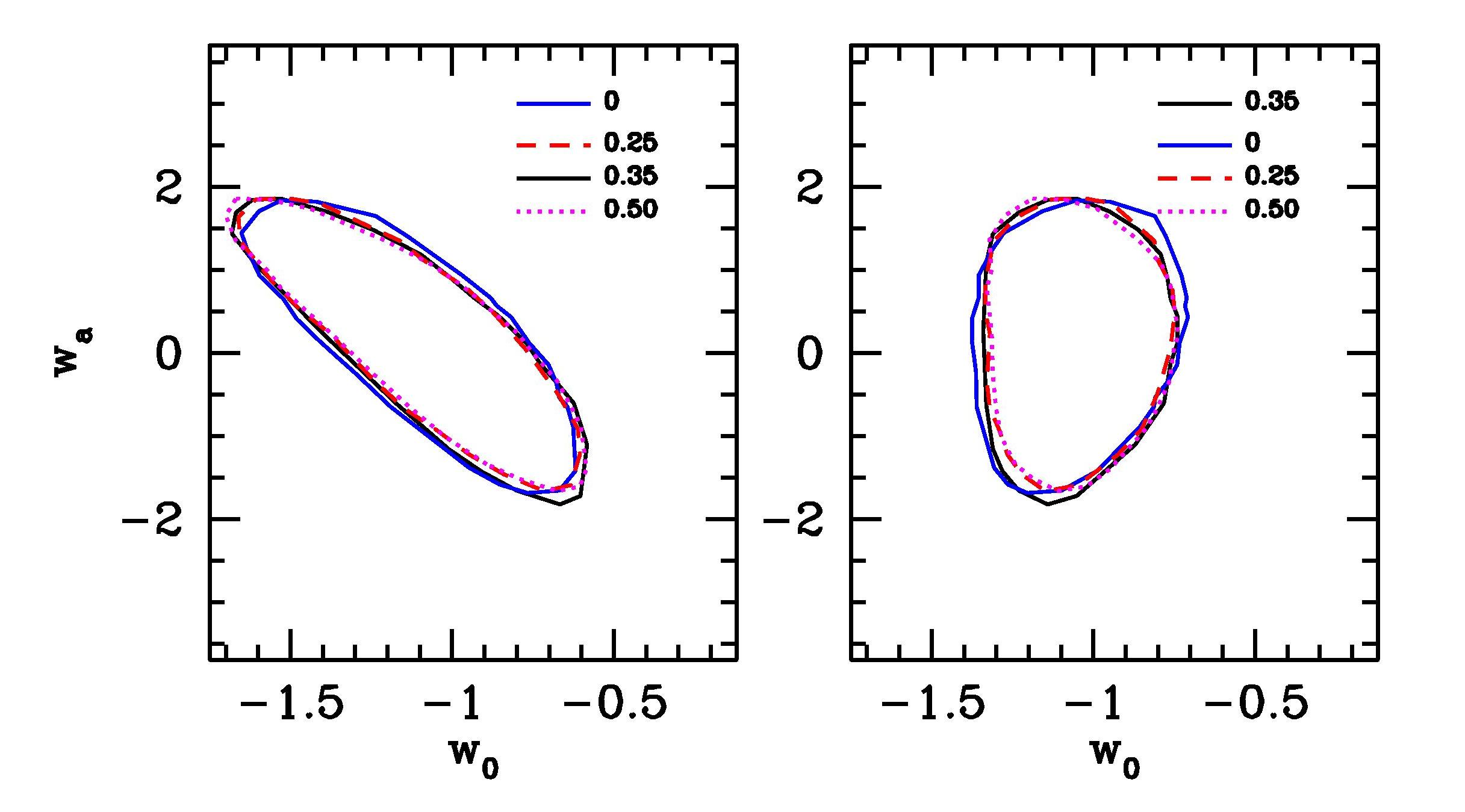}
\includegraphics[height=8cm,angle=0]{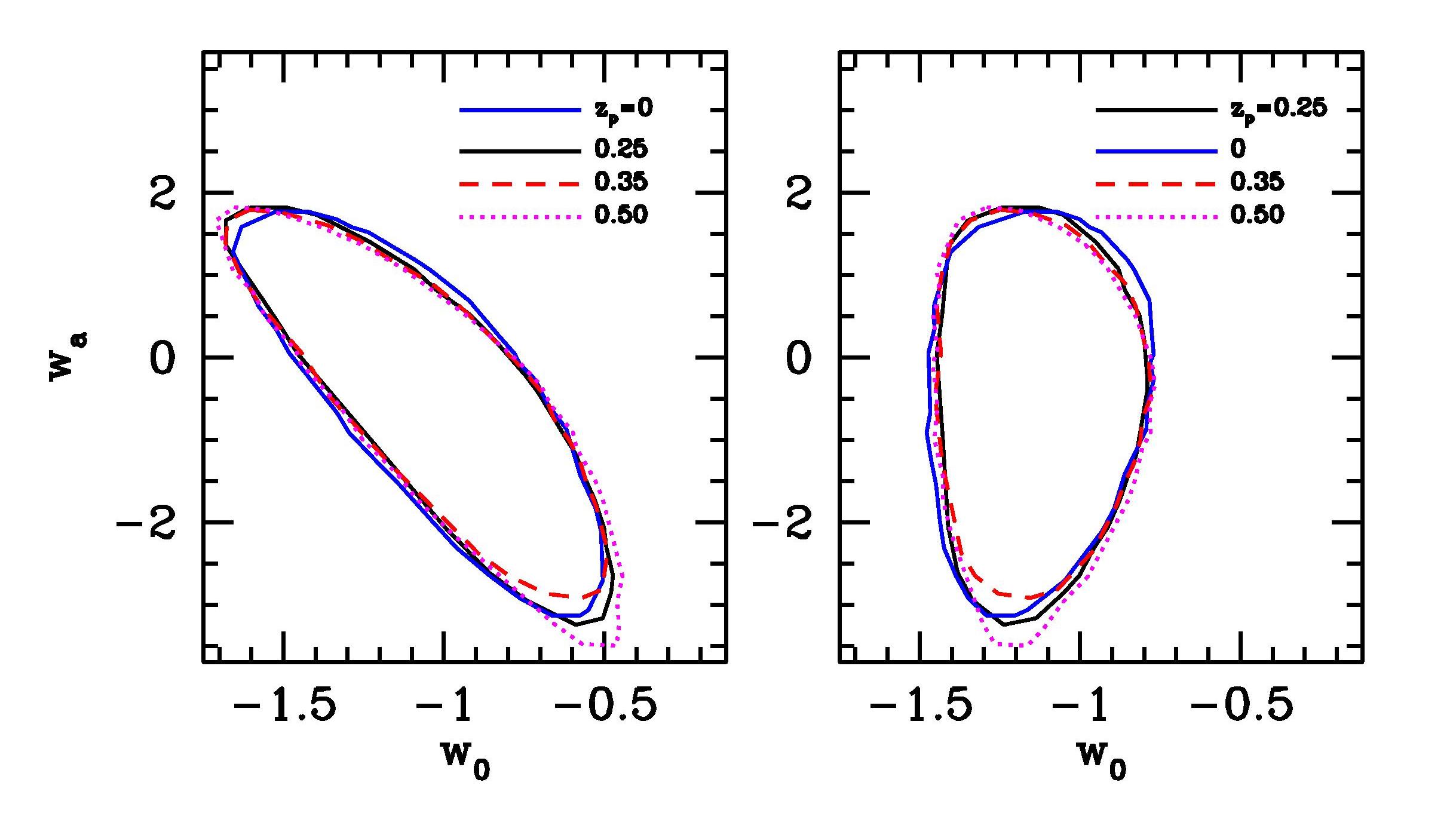}
\vskip -0.5truecm
\caption{2$\sigma$ marginalized likelihood contours on the $w_0-w_a$
  plane, obtained with different $z_p$ and translated to a unique
  redshift. All plots refers to CMB+SN+BAO+HST constraint combination.
  Plots at the l.h.s. (r.h.s.) are translated to $z=0$ (to the redshift
  where $w_0$ and $w_a$ constraints are uncorrelated). The 2 plots on
  the first (second) row are for $f_\nu \equiv 0$ ($f_\nu
  \neq 0).$
}
\label{fig5}
\vskip .5truecm
\end{figure}
%%%%%%%%%%%%%%%%%%%%%%%%%%%%%%%%%%%%%%%%%%%%%%%%%%%%%%%%%%%%%%%%%%%

These plots are one of the results of this analysis. They confirm the
high reliability of the MC algorithm, yielding close results when
different parameter combinations are fitted.

Residual slight differences between the plotted ellipses are a
measure of the reliability of the algorithm used.

%%%%%%%%%%%%%%%%%%%%%%%%%%%%%%%%%%%%%%%%%%%%%%%%%%%%%%%%%%%%%%%%%%%
\begin{table}
\vskip 1.1truecm
 \begin{tabular}{p{0.07\textwidth}|p{0.15\textwidth}p{0.15\textwidth}|p{0.15\textwidth}p{0.15\textwidth}p{0.17\textwidth}}

\hline
 & \hfill Massless& Neutrinos\hfill &\hfill Massive &Neutrinos\hfill&\\
 \hline
 \hline
$z_p$	&$w_0\pm1\sigma\pm2\sigma$        &$w_a\pm1\sigma\pm2\sigma$	& $w_0\pm1\sigma\pm2\sigma$ & $w_a\pm1\sigma\pm2\sigma$ & $f_\nu\pm1\sigma\pm2\sigma$   \\
\hline
\hline
$0$ & $-1.13_{-0.10-0.30}^{+0.09+0.33}$ & $0.29_{-0.25-1.30}^{+0.39+1.00}$&$-1.07_{-0.11-0.34}^{+0.10+0.38}$ &$-0.31_{-0.34-1.89}^{+0.56+1.37}$&$0.036_{-0.036-0.036}^{+0.009+0.041}$ \\ 
$0.1$ & $-1.11_{-0.07-0.22}^{+0.06+0.23}$     & $0.32_{-0.24-1.29}^{+0.40+0.97}$  &$-1.10_{-0.07-0.24}^{+0.07+0.24}$ &$-0.27_{-0.36-1.78}^{+0.53+1.34}$ &$0.035_{-0.035-0.035}^{+0.009+0.040}$\\
$0.2$         & $-1.08_{-0.05-0.16}^{+0.05+0.16}$     & $0.31_{-0.26-1.32}^{+0.40+1.03}$  &$-1.12_{-0.05-0.18}^{+0.05+0.18}$ &$-0.32_{-0.34-1.93}^{+0.58+1.36}$ & $0.036_{-0.036-0.036}^{+0.010+0.041}$   \\
$0.25$         & $-1.07_{-0.04-0.14}^{+0.04+0.14}$  & $0.32_{-0.23-1.27}^{+0.39+0.97}$ & $-1.13_{-0.04-0.18}^{+0.05+0.16}$ &$-0.28_{-0.36-1.95}^{+0.58+1.43}$ &$0.035_{-0.035-0.035}^{+0.009+0.041}$    \\
$0.35$         & $-1.05_{-0.03-0.14}^{+0.04+0.13}$  & $0.28_{-0.24-1.32}^{+0.39+0.99}$ & $-1.14_{-0.04-0.22}^{+0.06+0.17}$ &$-0.24_{-0.37-1.82}^{+0.55+1.36}$ &$0.035_{-0.035-0.035}^{+0.010+0.040}$    \\
$0.4$         & $-1.05_{-0.03-0.15}^{+0.04+0.13}$ & $0.29_{-0.24-1.33}^{+0.40+0.98}$ & $-1.15_{-0.04-0.26}^{+0.08+0.19}$ & $-0.30_{-0.36-1.86}^{+0.57+1.37}$ & $0.036_{-0.036-0.036}^{+0.009+0.041}$    \\
$0.5$         & $-1.03_{-0.03-0.18}^{+0.05+0.13}$  & $0.30_{-0.24-1.32}^{+0.40+0.96}$ & $-1.17_{-0.05-0.34}^{+0.10+0.22}$ &$-0.32_{-0.35-1.93}^{+0.57+0.40}$ &$0.035_{-0.035-0.035}^{+0.010+0.040}$    \\

\hline
\end{tabular}
\label{tavola2}
\caption{Mean and fully marginalized limits at $1\sigma$ and $2\sigma$ for different pivoting redshifts. In all the cases the data set is CMB+SN+BAO+HST.}
\end{table}

%%%%%%%%%%%%%%%%%%%%%%%%%%%%%%%%%%%%%%%%%%%%%%%%%%%%%%%%%%%%%%%%%%%
\begin{figure}
\centering
\includegraphics[height=8.8cm,angle=0]{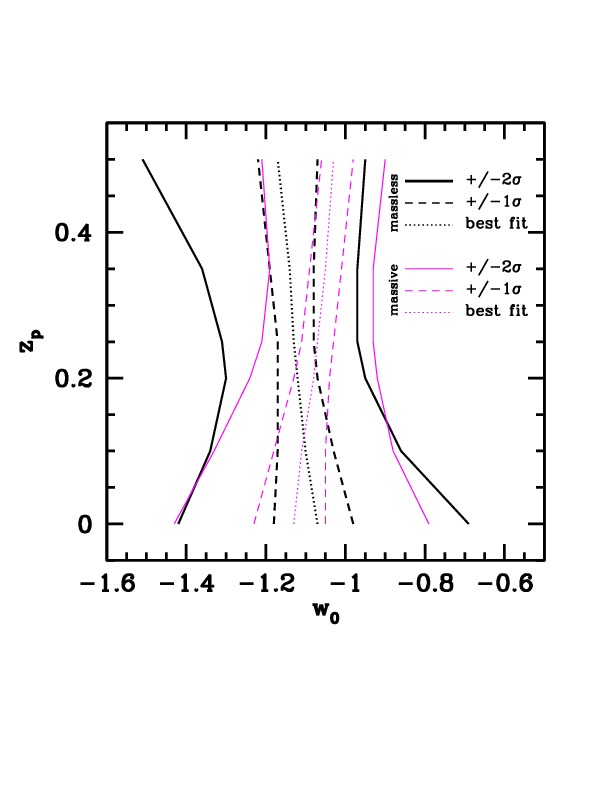}
\includegraphics[height=9.4cm,angle=0]{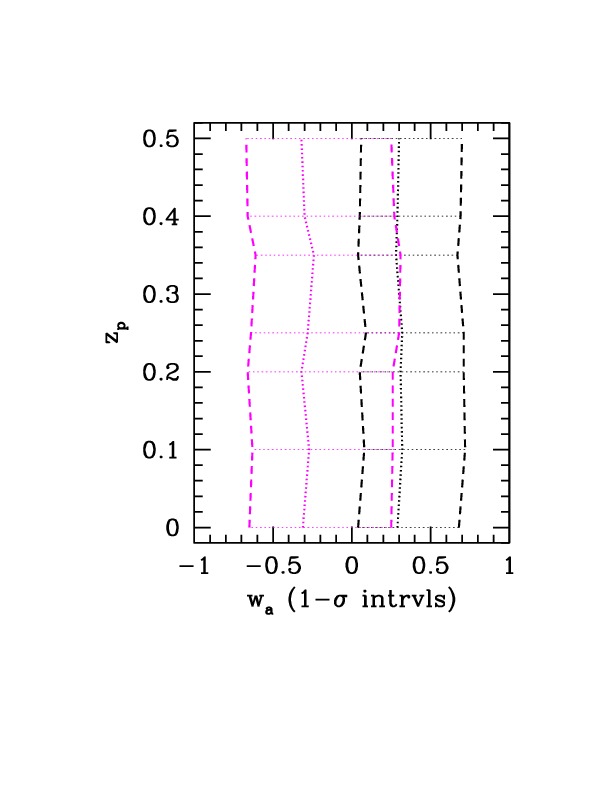}
\vskip -0.5truecm
\caption{On the l.h.s. (r.h.s) plot of mean, $1\sigma$ and
  $2\sigma$ limits on $w_0$ (mean and $1\sigma$ limits on $w_a$)
  at different $z_p$, for massless or massive neutrinos as indicated
  on the frame. In all cases, the constraint combination is
  (CMB+SN+BAO+HST).
}
\label{fig6}
\vskip .5truecm
\end{figure}
%%%%%%%%%%%%%%%%%%%%%%%%%%%%%%%%%%%%%%%%%%%%%%%%%%%%%%%%%%%%%%%%%%%

%\begin{figure}
%\centering
%\includegraphics[height=7cm,angle=0]{fnuwa_pivot.jpg}
%\includegraphics[height=11cm,angle=0]{fnuwa_pivot.ps}
%\vskip -.5truecm
%\caption{}
%\label{HM}
%\vskip -.5truecm
%\end{figure}
%%%%%%%%%%%%%%%%%%%%%%%%%%%%%%%%%%%%%%%%%%%%%%%%%%%%%%%%%%%%%%%%%%%

\section{Discussion}
Linear DE state equations $w(a)$ are expressed by using $w_{0,a_p}$
and $w_{a,a_p}$ coefficients (usually $w_{0}$ and $w_{a}$) which do
depend on the pivot $a_p$ value selected. Data yield constraints on
$w_{0,a_p}$ and $w_{a,a_p}$ which turn out to be uncorrelated for a
single $z_p$ value. We confirm that the pivoting redshift yielding no
correlation is $\simeq 0.35$, if $f_\nu = 0$, but we find that it
lowers to $\sim 0.25$ if a degree of freedom allowing for the
$\nu$--mass is open.

If we follow the procedure suggested by the Dark Energy Task Force
(\cite{detf}) to evaluate a Figure of Merit (FoM) for the precision of
the two fits, we find
\begin{equation}
[\sigma(w_a) \times \sigma(w_0)]^{-1} = 9.71~~{\rm or} ~~ 17.2
\end{equation}
in the $M_\nu \neq 0$ or $\equiv 0$ cases, respectively. The standard
deviations $\sigma(w_a)$ and $\sigma(w_0) $ are evaluated at the $z_p$
value allowing independent $w_a$ and $w_0$ estimates, by fitting a
Gaussian distribution on the posterior distribution. These values
agree with DETF expectations, suggesting a range between 6.1 and
35.2. By using WMAP5, BAO and SN data, \cite{wang} found FoM=8.3~. By
using just more recent BAO and SN constraints, \cite{SL} found
FoM=14.2. Our slightly greater FoM arises from the improved and wider
set of data. For the sake of comparison, still according to
\cite{wang1}, when linear laws are parametrized by the values of $w$
at $z=0$ and 0.5, the FoM is $\sim 25~$.

%%%%%%%%%%%%%%%%%%%%%%%%%%%%%%%%%%%%%%%%%%%%%%%%%%%%%%%%%%%%%%%%%%%
\begin{figure}
\centering
\includegraphics[height=8cm,angle=0]{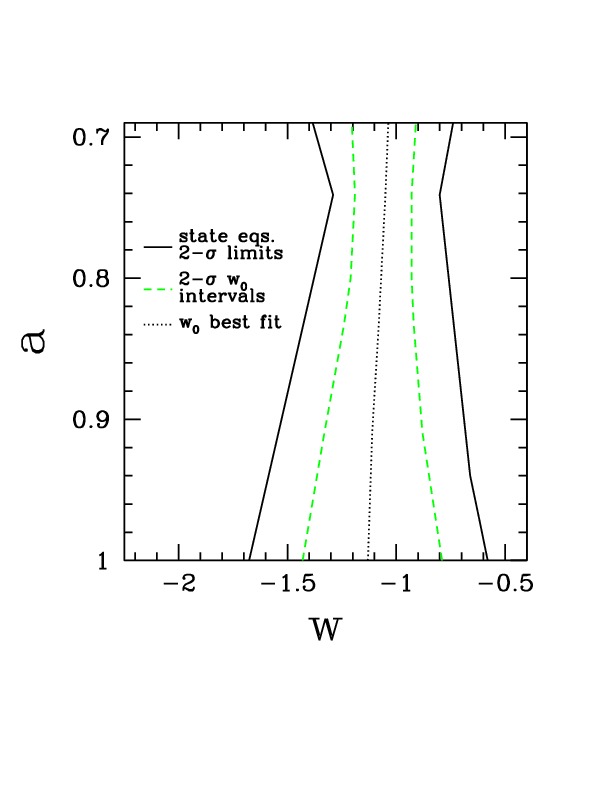}
\includegraphics[height=8cm,angle=0]{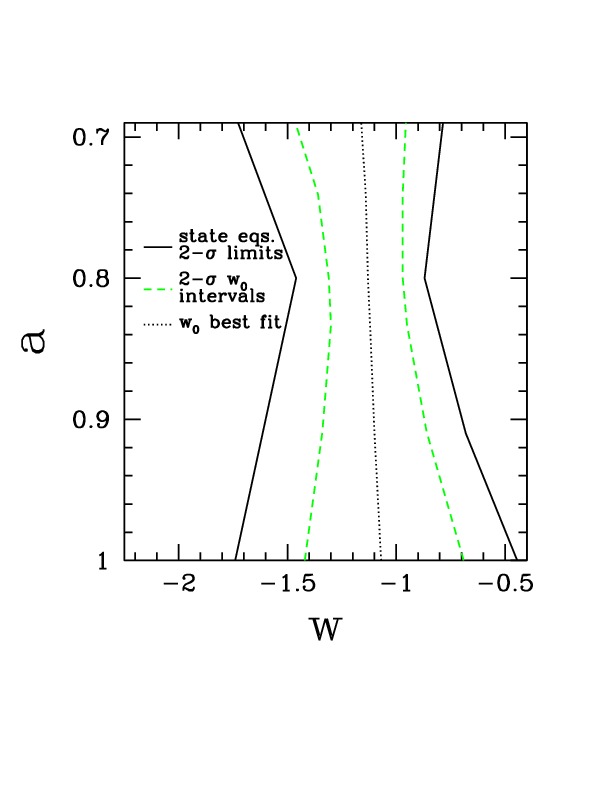}
\vskip -0.5truecm
\caption{Envelop of DE state equations (eq.~2). The plot is built by
  using $w_0$ and $w_a$ values with uncorrelated errors. Any straight
  line completely inside the black contours is an allowed DE state
  equation. Using other pivoting $z_p$ causes almost irrelevant
  changes (apart of the case $z_p = 0$ with $f_\nu = 0$). For the sake
  of comparison, the green dashed lines limit the $w_0$ interval
  allowed when we fit data by using varying $a_p$. The data set is
  (CMB+SN+BAO+HST).
}
\label{fig6}
\vskip .5truecm
\end{figure}
%%%%%%%%%%%%%%%%%%%%%%%%%%%%%%%%%%%%%%%%%%%%%%%%%%%%%%%%%%%%%%%%%%%

%%%%%%%%%%%%%%%%%%%%%%%%%%%%%%%%%%%%%%%%%%%%%%%%%%%%%%%%%%%%%%%%%%%

%%%%%%%%%%%%%%%%%%%%%%%%%%%%%%%%%%%%%%%%%%%%%%%%%%%%%%%%%%%%%%%%%%%

The $w_0$--$w_a$ ranges found in the cases $M_\nu \equiv 0$ and $M_\nu
\neq 0$ exhibit significant overlaps, as expected. Our finding is that
the allowed $w_0$ range, when marginalizing in respect to any other
parameter, exhibits different trends in the two cases: the $w_0$ range
tends to decrease (increase) when the option $f_\nu \neq 0$ is allowed
(disregarded). As a consequence, the $w_0$ intervals show no overlap
(at 1$\sigma$) when the pivot redshift exceeds $\sim 0.4$ (see
%Figure \ref{w0zpd}, l.h.s.). We also plot the $z_p$ dependence of
Figure 6, l.h.s.). We also plot the $z_p$ dependence of
%$w_a$ 1$\sigma$ errorbars (Figure \ref{w0zpd}, r.h.s.), confirming
$w_a$ 1$\sigma$ errorbars (Figure 6, r.h.s.), confirming
$w_a$ estimates to be smaller when $M_\nu \neq 0$ is allowed. Clearly,
no $z_p$ dependence is evident here. For more details results see
Table 2.

The set of linear $w(a)$ laws should however be independent of the
%pivoting redshift. In Figure \ref{m_vlp} we show the line envelops for
pivoting redshift. In Figure 7 we show the line envelops for
the cases $f_\nu = 0$ (l.h.s.) and $\neq 0$ (r.h.s.). In the same
plots we report also the $w_0$ constraints shown in the previous
Figure, and the best fit $w_0$ at any redshift. A hint of the trends
found in the evolution of the $w_0$ range (and $w_0$ best--fit value)
can be seen also in the average behaviors of the fitting linear laws.

The fact that the linear laws are the same, independently of the $z_p$
%chosen to perform the fit, is confirmed in Figures \ref{overlaps}.
chosen to perform the fit, is confirmed in Figures 5.
The plots confirm the good performance of the MC algorithm, yielding
overlapping results for different linear combinations of the fitting
parameters. 

\section{WMAP9 release}
After the completion of this work, nine--year WMAP data have been
released, together with a number of fits on cosmological parameters
(\cite{wmap9}). 
%As expected, these data yield (slightly) more stringent parameter estimates.

%Let us mention, in particular, the results on $w_0$,~$w_a$ parameters
%shown in Fig.~10 of the release. The ellipses colored in red are
%to be compared with the blue ellipses in Figure 1 (l.h.s.) of
%this work. (This is the only significant comparison allowed).

We had mentioned that our curves are displaced towards significantly
more negative $w_0$ and greater $w_a$ values, in respect to WMAP7
output analysis by \cite{wmap7}. WMAP9 ellipses fully confirm
such displacement. 

The main difference between ours and WMAP9's results concerns the
shape of the error distributions. Our peak about top likelihood values
appears more pronounced, with an error distribution becoming much
flatter above $\sim 1\sigma$. In fact, our 1$\sigma$ errors
apparently yield even tighter constraints than WMAP9: the $w_0$/$w_a$
intervals pass from -1.23,-1.04/0.04,0.68 (our values) to
-1.29,-1.04/-0.14,0.85 (WMAP9). The $w_0$ output are therefore
overlappable, while also $w_a$ is equally centered, with an error
drastically reduced from 0.99 to 0.64~.

The situation is clearly opposite if we consider the $w_0$ and $w_a$
intervals spanned by the 2$\sigma$ ellipse on the $w_0$--$w_a$
plane. By comparing Figure 10 in WMAP9 release with our Figure 1, we
see intervals passing from -1.51,-0.84/-1.12,1.41 to
-1.59,-0.61/-1.70,1.79 for $w_0$ and $w_a$, respectively.

The more pronounced Gaussian behavior in WMAP9 data is probably
related to their greater dataset, also implying an enhanced
disagreement for central and 1$\sigma$ values, but overall tighter
constraints.

%Once again $w_0$ outputs are overlappable; on
%the contrary, $w_a$ results share the interval -1.1,1.29, but comprise
%still smaller (greater) values, up to -1.59 (1.45) in our (WMAP9)
%case.
%% According to our
%Figure 1, the 2$\sigma$ curve in the $w_0$--$w_a$ plane spans the
%intervals -1.64/-0.62 and -1.88/1.84, respectively. The same
%intervals, in Figure 10 of the WAMP9 release are -1.5/-0.84 and
%-1.2/1.4, respectively. The interval width is then reduced from
%1.02--3.7 to 0.66--2.6 (for $w_0$ and $w_a$, respectively) with a
%significant improvement. The new ellipse, however, is completely
%comprised inside the one shown in Figure 1 and almost equally
%centered. In particular the shift, in respect to WMAP7 results, due to
%fresh SNIa data, is fully confirmed.

\acknowledgments

S.B. acknowledges the financial support of CIFS.



\begin{thebibliography}{99}
\bibitem[Damour et al.(1990)]{coupledquintessence1}T. Damour, Gibbons
  G. W., Gundlach C., Phys.Rev.Lett. 64 (1990) 123
\bibitem[Wetterich (1995)]{coupledquintessence2}C. Wetterich, A\&A 301
  (1995) 321; Amendola L., Phys.Rev.D 62 (2000) 643511
\bibitem[Amendola (2000)]{coupledquintessence3}Amendola L., Phys.Rev.D
  62 (2000) 643511
\bibitem[Amendola \& Quercellini (2001)]{coupledquintessence4}Amendola
  L., Quercellini C., Phys Rev D68 (2001) 023514
\bibitem[Dalal et al.(2001)]{coupledquintessence5}N. Dalal,
  K. Abazajian, E. E. Jenkins and A. V. Manohar, {\it Testing the
    cosmic coincidence problem and the nature of dark energy},
  Phys. Rev. Lett. 87 (2001), 141302
\bibitem[Amendola \& Tocchi Valentini (2002)]{coupledquintessence6}Amendola 
  L., Tocchi Valentini D., Phys. Rev. D66 (2002) 041528
\bibitem[Maccio' et al.(2004)]{coupledquintessence7bis}Andrea
  V. Maccio', Claudia Quercellini, Roberto Mainini, Luca Amendola and
  Silvio A. Bonometto, {\it N-body simulations for coupled dark
    energy: halo mass function and density profiles}, Phys.Rev.D69
  (2004) 123516
\bibitem[Mainini \& Bonometto (2004)]{coupledquintessence7} Roberto
  Mainini, Silvio A. Bonometto, {\it Dark Matter and Dark Energy from
    the solution of the strong CP problem}, Phys.Rev.Lett. 93 (2004)
  121301
\bibitem[Mainini \& Bonometto (2006)]{coupledquintessence8}Roberto
  Mainini, Silvio Bonometto, {\it Mass functions in coupled Dark
    Energy models}, Phys.Rev.D74 (2006) 043504
\bibitem[Mainini \& Bonometto (2007)]{coupledquintessence9}Mainini
  Roberto, Silvio Bonometto, {\it Dark Matter \& Dark Energy from a
    single scalar field: CMB spectrum and matter transfer function},
  JCAP 0709 (2007) 017
\bibitem[Bento \& Bortolami (2009)]{coupledquintessence10}M.C. Bento,
  O. Bertolami, {\it Dark energy and the Rutherford-Soddy radiative
    decay law} Phys.Lett.B675 (2009) 231
\bibitem[Bento \& Gonzales (2009)]{coupledquintessence11}M.C. Bento,
  R. Gonzalez Felipe, {\it The variation of the electromagnetic
    coupling and quintessence}, Phys.Lett.B674 (2009) 146-151
\bibitem[Bento et al.(2008)]{coupledquintessence12}M.C. Bento,
  R. Gonzalez Felipe, N.M.C. Santos {\it Brane assisted quintessential
    inflation with transient acceleration }, Phys.Rev. D77 (2008)
  123512
\bibitem[Zimdahl et al.(2001)]{coupledquintessence13}W. Zimdahl,
  D. Pavon, and L. P. Chimento, {\it Interacting quintessence},
  Phys.Lett.B 521 (2001) 133
\bibitem[Del Campo \& Herrera (2006)]{coupledquintessence14}S.~del
  Campo, R. Herrera, {\it G. Olivares, and D. Pavon, Interacting
    models of soft coincidence}, Phys.Rev.D 74 (2006) 023501
\bibitem[Wei \& Zhang (2007)]{coupledquintessence15} H. Wei and
  S. N. Zhang, {\it Observational H(z) data and cosmological models},
  Phys.Lett. B644 (2007) 7
\bibitem[Amendola et al.(2007)]{coupledquintessence16}L. Amendola,
  G. C. Campos, and R. Rosenfeld, {\it Consequences of dark
    matter-dark energy interaction on cosmological parameters derived
    from SN Ia data}, Phys.Rev.D 75 (2007) 083506
\bibitem[Guo et al.(2007)]{coupledquintessence17}Z. K. Guo, N. Ohta,
  and S. Tsujikawa, {\it Probing the coupling between dark components
    of the universe}, Phys.Rev.D 76 (2007) 023508
\bibitem[Caldera-Cabral et al.(2009)]{coupledquintessence18}
  G. Caldera-Cabral, R. Maartens and L. A. Urena-Lopez, {\it Dynamics
    of interacting dark energy}, Phys.Rev.D 79 (2009) 063518
\bibitem[Pettorino et al.(2012)]{coupledquintessence19}Valeria
  Pettorino, Luca Amendola, Carlo Baccigalupi, Claudia Quercellini,
  {\it Constraints on coupled dark energy using CMB data from WMAP and
    SPT}, arXiv:1207.3293
%
\bibitem[Capozziello et al.(2006)]{GRviolate1}S. Capozziello,
  S. Nojiri, S. D. Odintsov and A. Troisi, Phys. Lett. B 639 (2006)
  135
\bibitem[Amendola et al.(2007)]{GRviolate2}L. Amendola, D. Polarski,
  and S. Tsujikawa, {\it Are f(R) dark energy models cosmologically
    viable?} Phys.Rev.Lett. 98 (2007) 131302
\bibitem[Amendola et al.(2007)]{GRviolate3} L. Amendola, R. Gannouji,
  D. Polarski and S. Tsujikawa, Phys.Rev.D 75 (2007) 083504
\bibitem[Creminelli et al.(2009)]{GRviolate4}P. Creminelli,
  G. D'Amico, J. Norena, \& F. Vernizzi, {\it The Effective Theory of
    Quintessence: the w<-1 Side Unveiled}, JCAP 0902 (2009) 018
\bibitem[Park et al.(2010)]{GRviolate5} M. Park, K. M. Zurek and
  S. Watson, {\it A Unified Approach to Cosmic Acceleration},
  Phys.Rev.D 81 (2010) 124008
\bibitem[Bloomfield \& Flanagan (2012)]{GRviolate6} J. K. Bloomfield
  and E. E. Flanagan, {\it A Class of Effective Field Theory Models of
    Cosmic Acceleration}, JCAP 10 (2012) 039
%
\bibitem[Tomita (2000)]{void1}K. Tomita, {\it Distances and lensing in
  cosmological void models}, Astrophys.J. 529 (2000) 38
\bibitem[Celerier (2000)]{void2}M. N. Celerier, {\it Do we really see
  a cosmological constant in the supernovae data?}, A\&A 353 (2000)
  63
\bibitem[Tomita (2001)]{void3}K. Tomita, {\it A local void and the
  accelerating universe}, MNRAS 326 (2001) 287
\bibitem[Iguchi et al.(2002)]{void4}H. Iguchi, T. Nakamura, and
  K. i. Nakao, {\it Is dark energy the only solution to the apparent
    acceleration of the present universe?}, Prog.Theor.Phys. 108
  (2002) 809
\bibitem[Jimenez et al. (2012)]{void5}R. Jimenez, P. Talavera,
  L. Verde, {\it An effective theory of accelerated expansion},
  arXiv:1107.2542
%
\bibitem[Colombo et al.(2009)]{coupledfit1}L.P.L. Colombo, R. Mainini and
  S.A. Bonometto, {\it Do WMAP data favor neutrino mass and a
    coupling between Cold Dark Matter and Dark Energy?},  G. La Vacca,
  J.R. Kristiansen, JCAP 0904 (2009) 007
\bibitem[Mainini (2009)]{coupledfit2}Roberto Mainini, {\it Voids and
  overdensities of coupled Dark Energy}, JCAP 0904 (2009) 017
\bibitem[Kristiansen et al.(2010)]{coupledfit3}J. R. Kristiansen,
  G. La Vacca, L. P. L. Colombo, R. Mainini, S. A. Bonometto, {\it
    Coupling between cold dark matter and dark energy from neutrino
    mass experiments}, NewAstron.15 (2010) 609
\bibitem[Chevallier \& Polarski (2001)]{chevall} Chevallier M. \&
  Polarski D., Int.J.Mod.Phys. D10 (2001) 213
\bibitem[Wang (2008)]{wang} Wang Y., {\it Figure of Merit for Dark
  Energy Constraints from Current Observational Data}, Phys.Rev.D 77 (2008)
  123525
\bibitem[Salzano et al. (2012)]{wang1} Salzano V., Wang Y., Sendra I.,
  Lazkoz R., {\it Linear dark energy equation of state revealed by
    supernovae?}, arXiv:1211.1012 (2012)
\bibitem[Sendra \& Lazkoz (2012)]{SL} Sendra I., Lazkoz R., {\it SN
  and BAO constraints on (new) polymonial dark energy parametrization:
  current results and forecasts}, MNRAS 422 (2012) 776

%
%%%%algoritmo e dati%%%%%%%%%%
\bibitem[Lewis et al.(2002)]{cosmomc}Lewis A., Bridle S., {\it
  Cosmological parameters from CMB and other data: A Monte Carlo
  approach}, Phys.Rev.D {\bf 66}  (2002) 103511
%
\bibitem[Fang et al.(2008)]{ppf}Fang W., Wang S., Hu W., Haiman Z.,
  Hui L., May M., {\it Challenges to the DGP model from horizon--scale
    growth and geometry}, Phys.Rev.D {\bf 78} (2008) 103509
%
\bibitem[Suzuki et al.(2012)]{union21}Suzuki N. {\it et al.}, {\it The
  Hubble Space Telescope Cluster Supernova Survey: V. Improving the
  Dark Energy Constraints Above z>1 and Building an Early-Type-Hosted
  Supernova Sample,} arXiv:1105.3470, ApJ {\bf 746} (2012) 85
%
\bibitem[Blake et al.(2011)]{wigglez}Blake {\it et al.}, {\it The
  WiggleZ Dark Energy Survey: measuring the cosmic expansion history
  using the Alcock-Paczynski test and distant supernovae},
  arXiv:1108.2637 
%
\bibitem[Percival et al.(2010)]{sdss}Percival {\it et al.}, {\it
  Baryon Acoustic Oscillations in the Sloan Digital Sky Survey Data
  Release 7 Galaxy Sample}, Mon.Not.Roy.Astron.Soc. {\bf 401},
  2148-2168, (2010), arXiv:0907.1660
%
\bibitem[Riess et al.(2009)]{hst}Riess {\it et al.}, {\it A
  redetermination of the Hubble constant with the Hubble Space
  telescope from a differential distance ladder} ApJ. {\bf 699} (2009)
  539-563
%
\bibitem[Lewis et al.(2000)]{camb}Lewis A., Challinor A. , Lasenby A.,
  {\it Efficient Computation of CMB anisotropies in closed FRW
    models}, arXiv:astro-ph/9911177, ApJ {\bf 538} (2000) 473 
%
\bibitem[Komatsu et al.(2011)]{wmap7}Komatsu E. {\it et al.}, {\it
  Seven year Wilkinson microwave anisotropy probe (WMAP) observations:
  cosmological interpretation}, arXiv:1001.4538, ApJ.Suppl. {\bf 192}
  (2011) 18
%
\bibitem[De Bernardis et al. (2008)]{phantom} De Bernardis F., Serra
  P., Cooray A., Melchiorri A., {\it An improved limit on the neutrino
    mass with CMB and redshift-dependent halo bias-mass relations from
    SDSS, DEEP2, and Lyman-Break Galaxies}, Phys.Rev. D78 (2008) 083535
    
\bibitem[Albrecht et al. (2006)]{detf} Albrecht A., Bernstein G., Cahn
  R., Freedman W. L., Hewitt J., Hu W., Huth J., Kamionkowski M., Kolb
  E.W., Knox L., Mather J.C., Staggs S., Suntzeff N.B., {\it Report of
    the Dark Energy Task Force}, arXiv:astroph/0609591 (2006)
%	
\bibitem[Hinshaw et al.(2012)]{wmap9}Hinshaw G. {\it et al.}, {\it
  Nine year Wilkinson microwave anisotropy probe (WMAP) observations:
  cosmological parameter results}, arXiv:1001.4758

\end{thebibliography}
\end{document}